# ON THE INFLUENCE OF EXPLAINABLE AI ON AUTOMATION BIAS

*Research in Progress*

Max Schemmer, Karlsruhe Institute of Technology, Germany, max.schemmer@kit.edu

Niklas Kühl, Karlsruhe Institute of Technology, Germany, niklas.kuehl@kit.edu

Carina Benz, Karlsruhe Institute of Technology, Germany, carina.benz@kit.edu

Gerhard Satzger, Karlsruhe Institute of Technology, Germany, gerhard.satzger@kit.edu

## Abstract

*Artificial intelligence (AI) is gaining momentum, and its importance for the future of work in many areas, such as medicine and banking, is continuously rising. However, insights on the effective collaboration of humans and AI are still rare. Typically, AI supports humans in decision-making by addressing human limitations. However, it may also evoke human bias, especially in the form of automation bias as an over-reliance on AI advice. We aim to shed light on the potential to influence automation bias by explainable AI (XAI). In this pre-test, we derive a research model and describe our study design. Subsequentially, we conduct an online experiment with regard to hotel review classifications and discuss first results. We expect our research to contribute to the design and development of safe hybrid intelligence systems.*

## 1    Introduction

Over the last years, artificial intelligence (AI) has made significant progress (Lecun et al. 2015) and is being more and more integrated into standard applications and, thus, is becoming commoditized. However, especially in high-stake decision-making environments—such as medicine or banking—fully autonomous AI may be neither desirable nor (legally) possible. Additionally, AI capabilities are often limited to narrowly defined application contexts as the utilized algorithms often struggle to handle instances that differ from the patterns learned during training (D'Amour et al. 2020). In these cases, humans can contribute and leverage their unique capabilities such as intuition, creativity, and common sense. This line of thought gives rise to the vision of so-called hybrid intelligence (Dellermann et al. 2019), proposing to combine the complementary capabilities of humans and AI.

While AI has many advantages (like the capability to process significant data volumes), it may evoke human bias that negatively affects the outcome of hybrid intelligence (Challen et al. 2019). Specifically, automation bias (AB) is the "tendency to use automated cues as a heuristic replacement for vigilant information seeking and processing" (Mosier and Skitka 1999)—essentially representing an over-reliance on AI advice. Automation bias may trigger severe consequences as illustrated by human drivers directing their cars into rivers when blindly adopting erroneous routing advice (Santiago 2019). Similarly, Alberdi et al. (Alberdi et al. 2004) report that oncologists interpreting mammograms are accepting incorrect system advice in 33% to 40% of all cases—with serious impacts for their patients.

Research on AB in AI-based decision-making is still rare in IS research (Jussupow et al. 2021). This is plausible for two reasons: First, in the nascent, rather explorative phase of AI with often low AI performance, under-reliance was the more prominent issue. Second, prior work in IS often just assumed system advice to be perfectly correct (Jussupow et al. 2021). Thus, humans should always agree with the AI, and an issue of over-reliance could not arise per definition. However, with the increasing commoditization of AI and its application in more high-stake areas both have upped its performance as well as the realization of its imperfection. So, both understanding AB and design mechanisms to avoid it enjoy increasing research interest and practical relevance (Jussupow et al. 2021; Liel and Zalmanson 2020).

To address AB, humans should be enabled to have a basic understanding of the data, algorithms, and capabilities that determine the AI advice (Dzindolet et al. 2003). These insights would help them to identify incorrect AI advice (Arora 2020). The emerging research stream of explainable AI (XAI) (Bauer et al. 2021) aims to equip human users with such insight into AI decisions. This might help the human decision-maker to better judge the quality of an AI advice. An analogy would be the interaction between a management consultant, who provides advice, and his client as the decision-maker. To assess the quality of the recommendation, the decision-maker will ask the consultant to describe the reasoning. Based on the explanations, the decision-maker should be able to determine whether the advice can be relied upon or not. The same logic should hold for an "AI consultant". Explanations, thus, can also be seen as discrimination support for the decision-maker to assess whether to rely on the AI.

On the other hand, experimental studies indicate that explanations in human-AI collaboration can lead to "blind trust", i.e. following the AI advice without any evaluation of its trustworthiness (Bansal et al. 2021; Zhang et al. 2020). Research shows that explanations are sometimes interpreted more as a general sign of competence (Buçinca et al. 2021) and have a persuasive character (Bussone et al. 2015). This is supported by psychology literature which has shown that human explanations cause listeners to agree even in the presence of wrong explanations (Koehler 1991). The same effect could occur when the explanations are generated by AI. This potential effect of explanations on AB has severe consequences for the design of human-AI systems, especially if status-quo explanations are mostly seen as meaningful.

The ambiguous tradeoff between enabling the humans to discriminate the AI's advice, thereby reducing AB, and on the other hand, the tendency that the sole existence of an explanation could increase the reliance of the human on the AI which increases AB needs to be studied thoroughly. We, therefore, intend to shed light on the opposing effects that XAI may exert on AB by asking:

> **RQ:** *How does explainable artificial intelligence affect automation bias?*

To answer this question, we propose a research model, including five hypotheses, and design an online experiment to test our hypotheses. In a between-subject study, participants classify whether hotel reviews are genuine or fake, with the support of an AI—once without and once with explanations. First results of our pre-test with a limited number of participants indicate that explanations tend to increase AB. In the next step of our work, we intend to run our main study with sufficient participants to derive more insights. While certainly concrete tradeoffs of explanation effects on AB hinge on the domain and type of explanations, even our preliminary results hint towards the popular (and often undisputed) call that XAI may come at a price—the over-reliance on AI-based decisions.

With our work, we take a new perspective on XAI affecting automation bias. By providing initial insights into XAI and AB, we contribute to advancing the research field of hybrid intelligence. We expect our work to have concrete managerial implications informing the design of hybrid intelligence systems, especially for critical scenarios, such as medical diagnoses.

## 2      Related Work and Foundations

In the following, we briefly present the foundations of our work hybrid intelligence, AB, and XAI and subsequentially outline related work.

Dellermann et al. (2019) define **hybrid intelligence** as the ability to achieve goals by pairing humans and AI, thereby reaching superior results to those each of them could have accomplished individually.

Generally, the idea of hybrid intelligence is the combination of different, but complementary capabilities. While humans can rely on their senses, perceptions, emotional intelligence, and social skills (Braga and Logan 2017), AI excels at detecting patterns or calculating probabilities (Dellermann et al. 2019). These different skill set should in theory allow for superior complementary team performance in specific tasks (Hemmer et al. 2021).

Hybrid intelligence is dampened if **AB** prevents an effective human-AI collaboration. Traditionally, AB was studied in aviation automation. However, it can also be explicitly applied to human-AI collaboration (Challen et al. 2019). In their recent work, Jussupow et al. (2021) highlight the gap of AB coverage in IS research since most prior work has assumed that system advice is correct and, therefore, AB would not be an issue.

Goddard et al. (2012) state that the most rigorous form to quantitatively measure AB is the proportion of *negative consultations*, i.e., switches from a correct human decision to an erroneous one after incorrect AI advice. Negative consultations can be further distinguished into omission and commission errors—traditionally defined for binary monitoring tasks characterized by a default state (e.g., a patient is healthy) and a state of interest (e.g., the patient has a disease). *Omission errors* refer to instances where decision-makers fail to act despite their initial opinion because the AI did not indicate the state of interest (Skitka et al. 1999). An omission error in disease detection would mean that the AI fails to detect the disease and leads the doctor to not further pursue his initial diagnosis (Lyell and Coiera 2017). *Commission errors*, on the other hand, refer to cases in which humans follow flawed AI recommendations (Skitka et al. 1999): A doctor may act on the erroneous recommendation—e.g., believe the patient to have cancer and make him undergo chemotherapy or surgery. To measure negative consultations, an empirical study needs a sequential human-AI collaboration setup with two steps of human decision-making. Table 1 gives an overview of the different metrics based on a binary decision task. Note that for the sake of simplicity we use a binary decision task, but this could also be extended to multi-class and regression problems. Ground truth refers hereby to the correct decision, which needs to be known for the experimental setup. The sequential decision process can be described as follows: Initially, the human makes a decision, then he receives AI advice before finally the human is asked to update his decision. Other outcomes of sequential human-AI collaboration are possible, for example, in cases where all three choices are identical and correct. However, in this work, we focus on negative consultations as AB-related metrics.

| Ground Truth | Human choice before AI advice | AI advice | Human choice after AI advice | Metric | |
|---|---|---|---|---|---|
| 1 | 1 | 0 | 0 | Negative consultation | Omission error |
| 0 | 0 | 1 | 1 | | Commission error |

*Table 1.    Metrics of automation bias for a binary decision task.*

**Explainability** is a concept with a long tradition in Information Systems (IS) (Gregor and Benbasat 1999; Khasawneh and Kornreich 2014). With the rise of knowledge-based systems, expert systems, and intelligent agents in the 1980s and 1990s, the IS community laid the basis for research on explainability (Meske et al. 2020). The term XAI was first used by Van Lent et al. (2004) to describe the ability of their system to explain the behavior of agents. The current rise of XAI is driven by the need to improve the interpretability of complex models (Wanner et al. 2020). In comparison to interpretable linear models, more complex models can achieve higher performance (Briscoe and Feldman 2011). However, their inner workings are hard to grasp for humans.

XAI encompasses a wide spectrum of algorithms. In general, they can be differentiated by their complexity, their scope, and their level of dependency (Adadi and Berrada 2018). Wanner et al. (2020) cluster different types of complexity in white, grey, and black-box models. They define white-box models as models with perfect transparency, such as linear regressions. These models do not need additional explainability techniques but are intrinsically explainable. Black-box models, on the other hand, tend to achieve higher performance but lack interpretability. Lastly, grey-box models are not inherently interpretable but are made interpretable with the help of additional explanation techniques. These techniques can be differentiated in terms of their scope, i.e., being global or local explanations (Adadi and Berrada 2018). Global XAI techniques address holistic explanations of the models as a whole. In contrast, local explanations work on an individual instance basis. Besides the scope, XAI techniques can also be differentiated with regards to being model specific or model agnostic, i.e., can be used with all kinds of models (Adadi and Berrada 2018).

In terms of **related work** on XAI and AB, the studies of Buçinca et al., 2021 and Nourani et al., 2021 represent relevant work: Buçinca et al., 2021 investigate the impact of cognitive forcing functions to reduce over-reliance on XAI. Nourani et al., 2021 specifically investigate the anchoring effect and its relationship to AB. However, to the best of our knowledge, there exists no current research that studies how XAI influences AB by deriving and testing a research model.

# 3   Research Model

This pre-test aims to investigate the influence of XAI on AB. Accordingly, the dependent variables in our research model are measures of AB, i.e., negative consultations in general and omission and commission errors in particular. AB essentially describes the situation where a human would alone be able to solve a task but receives erroneous AI advice and follows this advice.

In our research, we not only want to analyze the direct impact of XAI on AB but additionally investigate the effect of mediators. Therefore, we systematically identify constructs from the literature that should influence AB. This approach highlights two important constructs that have historically shown an empirically strong effect on AB and are also influenced by XAI—trust and the human mental model (MM) (Bansal et al. 2021; Goddard et al. 2012). Trust is best conceptualized as an attitude towards AI (Lee and See 2004). Trust guides human decision-making when complexity makes a complete understanding of the AI not possible (Lee and See 2004). The concept of MMs has its roots in psychology (Craik 1943) and is characterized by the phenomenon that humans build small-scale models of how the world works. MMs are "representations of objects, events, and processes that people construct through interaction with their environment" (Savage-Knepshield 2001, p. 2). MMs enable individuals to understand a system's behaviors in terms of its "purpose (why the system exists)", "function (how the system operates)", "state (what the system is doing)", and "form (what the system looks like)" (Rouse and Morris 1986, p. 351). People create MMs for any system they interact with, including AI (Kulesza et al. 2012). In the following, we discuss the influence of XAI on the two mediators, trust and MM, and subsequentially their influence on AB (the resulting hypotheses and their relationships are depicted in Figure 1).

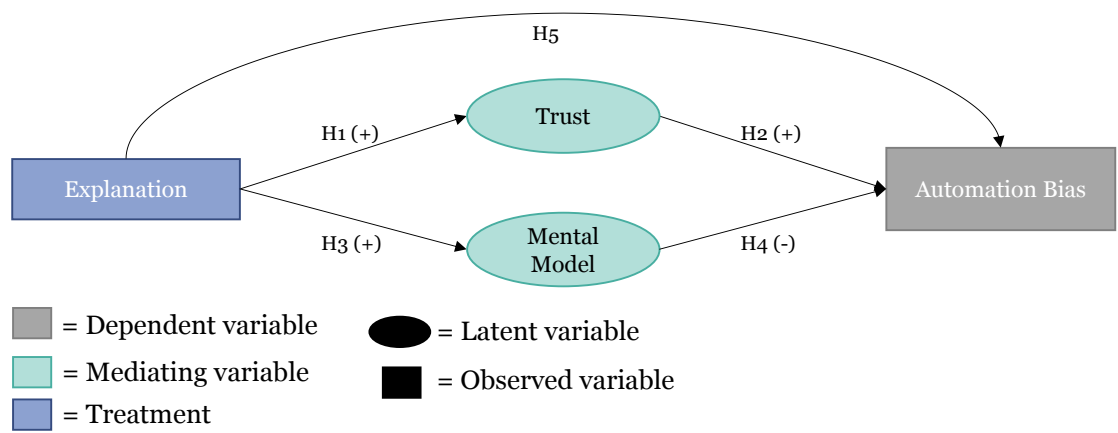

*Figure 1.      Research Model*

In recent work, Bansal et al. (2020) found that some explanations increased human's "blind" trust rather than calibrated it in a meaningful way. Psychological research shows that in general, explanations of humans increase trust (Koehler 1991). Sometimes explanations are more interpreted as a general sign of competence and thereby increase trust (Buçinca et al. 2021). Similar, research on human-AI interaction showed an increase in trust due to XAI (Weitz et al. 2019). We therefore hypothesize:

**H1:** *Providing explanations of the AI's reasoning increases trust in the AI.*

Trust is one of the most substantial reasons for AB (Goddard et al. 2012) as indicated by research in aviation automation that shows that humans with more trust are more likely to over-rely (Bailey and Scerbo 2007). In theory, trust should be calibrated and match the AI's capabilities (Lee and See 2004). Too low trust is called distrust and if the trust exceeds the capabilities it is called overtrust (Lee and See 2004). The most extreme level of trust might be "blind" trust, where the human does not evaluate the trustworthiness of the AI anymore (Bansal et al. 2020). While there is a tendency that trust in general will influence AB, especially overtrust and "blind" trust should influence AB. Thus, we hypothesize:

**H2:** *Increasing trust in the AI increases automation bias.*

Effective development of a MM includes the AI's background knowledge, capabilities, and preferences (Grosz 2012). AI explanations provide humans with indirect information on the AI's capabilities (Bansal et al. 2019) and should, therefore, improve the MM (Weld and Bansal 2019). Consequently, we hypothesize:

**H3:** *Providing explanations of AI's reasoning positively influences the reported human mental model of the AI.*

A recent paper by Hoffman et al. (2018) hypothesizes that forming a good MM of the AI is key to appropriate usage of AI. A MM of the AI model essentially represents an understanding of the AI's error boundaries, i.e., knowing when the AI will err (Bansal et al. 2019). Humans can then use this understanding of the error boundaries to contrast it with their own perceived capabilities and should thereby be able to detect erroneous advice. This is one possibility to reduce AB, as humans then simply reject the detected erroneous advice. An improved MM should improve the detection of erroneous AI advice. Thus, we hypothesize:

**H4:** *Positively influencing the reported human mental model of the AI decreases automation bias.*

Regarding the total effect of XAI on AB, we see several possible opposing effects. First, research on XAI has often hypothesized that explanations, in general, should improve the case-by-case evaluation of AI advice leading to an appropriate reliance on AI (Bansal et al. 2021; Schemmer et al. 2022; Zhang et al. 2020). For example, Bansal et al. (2021) hypothesize that if explanations do not "make sense", humans will reject the AI advice. Additionally, explanations should improve the MM (H3) and thereby further increase the potential of detecting erroneous AI advice. Rejecting erroneous AI advice, consequentially, reduces AB. However, on the other hand, explanations might also increase trust in AI (H1) which might result in an increase of AB (H2). Which effect exceeds the other is unclear. Thus, we hypothesize without the direction the total effect of XAI on AB:

**H5:** *Providing explanations of the AI's reasoning influences automation bias.*

We test the research model in an online experiment, which we describe in the following section.

## 4 Experimental Design

The research model is tested in an online experiment with a between-subject design. As an experimental task, we have chosen a deceptive hotel review classification task that is widely accepted in XAI research (Lai et al. 2020; Liu et al. 2021). Ott et al. (2011, 2013) provide the research community with a data set of 400 deceptive and 400 genuine hotel reviews. The deceptive ones were created by crowd-workers through MTurk. This setup allows for a clear ground truth. The dataset is highly used and well accepted in AI and specifically in deceptive review research (Lai et al. 2020).

We test two conditions, one in which the human is provided with the mere AI recommendation and one in which additional explanations are provided. The implemented AI is based on a Support Vector Machine with an accuracy of 86%, which is a performance that is similar to the performance in related literature (Lai et al. 2020). For the XAI condition, we use a state-of-the-art explanation technique, namely LIME feature importance explanations (Ribeiro et al. 2016). Feature importance means that the influence of a variable on the AI's decision is displayed with a numerical value. Since we are working with text data, a common technique to display the values is to highlight the respective words (Lai and Tan 2019). We additionally provide information on the direction of the effect and differentiate the values into three effect sizes following the implementation of (Lai et al. 2020; Lai and Tan 2019; Liu et al. 2021) (cf. Figure 2).

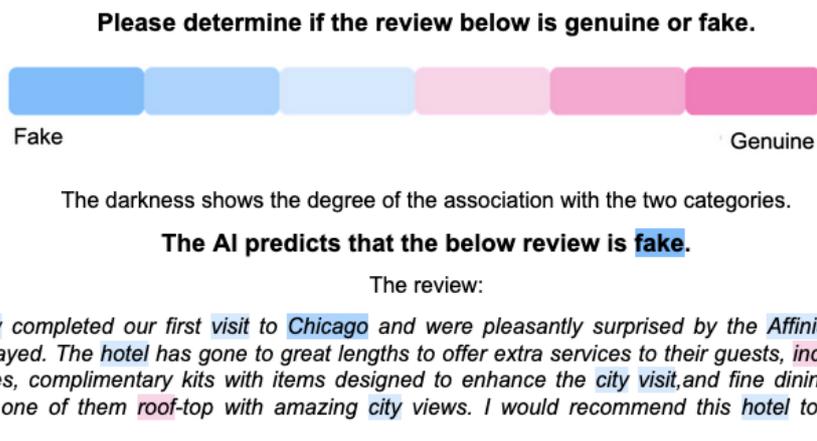

Figure 2.   Online experiment graphical user interface for the XAI treatment. The ground truth of the displayed hotel review is "fake".

In each treatment, participants are provided with 16 reviews. We incorporate an advanced sampling strategy to ensure a significant number of possible negative consultations. We have a test set of 32 reviews to which we apply stratified sampling and select four reviews of each class of a confusion matrix (True Positive, False Positive, True Negative, False Negative), two with a positive sentiment and two with a negative sentiment. This approach allows us to ensure a high-performing AI that should provide good explanations but also the potential for AB since half of the reviews provide wrong recommendations.

The participants for this study are recruited using the platform Prolific.co. The online experiment is initiated with an attention control question that asked participants to state the color of grass. To control for internal validity, participants are randomly assigned to the treatment groups, as suggested by Loewenstein (1999). Afterward, both treatment groups get an introduction to the task. Then, the participants conduct two training tasks to familiarize them with the task and the AI and, depending on the treatment, with its explanations. Additionally, the participants receive feedback on the training tasks. After the two training tasks, the participants are provided with the 16 main tasks. The participants did not receive feedback between the tasks on their performance or the ground truth of the reviews since we were interested in the isolated effect of XAI on AB. To incentivize the participants, they were informed that for every correct decision, they get an additional 12 Cent. Hereby, the two training classifications do not count for the final evaluation. The base payment is 5.83 Euro. After classifying the hotel reviews, we collect data on trust, MM, and demographic variables (age, gender, AI experience[1]). Lastly, we asked them an open question regarding their usage of the AI advice.

Trust is measured with the validated construct of Gefen et al. (2003). We exclude two items that are not applicable for our use case as they are specifically targeted toward online shopping. As a measurement for MM, we relate to the validated subconstructs of Vitharana et al. (2016)—goal, process, task, and information. In this pre-test, we focus on the sub-construct MM goal. All items are measured on a seven-point Likert scale.

---

[1] Measured with the following item: "How would you describe your knowledge in the field of Artificial Intelligence?"

# 5 Preliminary Results

In this research-in-progress article, we report qualitative feedback from the experiment participants and quantitative differences between the AI and XAI treatment from the results of our pre-test. Additionally, we perform a power analysis to calculate a valid sample size for the main study. To run the pre-test, we recruit 44 participants. We implement three attention checks and filter out all participants that fail more than one attention check, which leads to a final data set of 41 participants (21 for the XAI treatment and 20 for the AI treatment).

In a first step, we analyze the qualitative feedback of our study. In total, we received six comments (3 AI, 3 XAI). In the AI treatment, two comments are clearly positive towards the study setup and state they find the task interesting and have no issues. One comment raises an interesting aspect: *"I felt troubled with the reviews that I thought were genuine, but the AI told me were fake"*. This comment indicates that there may be differences in how the two different recommendations (fake or genuine) are perceived by participants. In the XAI setting, the participants commented three times on their uncertainty in their decision process: *"My decisions were based on gut reactions to the texts presented rather than any real insight into whether the reviews were genuine or fake. I felt like I had limited information to make an informed decision and the AI's input seemed arbitrary and not trustworthy."* These statements indicate that the participants might have problems understanding the explanations of the AI.

Next, we calculate the proportion of negative consultations, omission, and commission errors over all cases where the participants received incorrect AI advice. For the two constructs, we calculate the average of the respective items. We then test for significance with a two-sided t-test for the two mediators and a Mann-Whitney U test for negative consultations, omission, and commission errors since the assumption of a normal distribution is rejected for the AB measures. It must be noted that our data set for this pre-test is limited and only provides first insights. Table 2 visualizes our results.

In terms of behavioral constructs, trust does not seem to differ between both settings ($M_{AI}$ = 4.44, $M_{XAI}$ = 4.37) (**H1**). However, our pre-test indicates a difference of the MM goal ($M_{AI}$ = 5.1, $M_{XAI}$ = 4.37), i.e., participants in the AI treatment indicated a higher understanding of the goals of the AI (**H2**). In terms of **H3** and **H4**, we evaluated the Spearman correlation between trust and negative consultations ($r(42) = 0.13$, $p = 0.42$) and the MM goal and negative consultations ($r(42) = -0.14$, $p = 0.4$). Current results are not significant. However, we can see a trend in terms of the direction of the correlation which is in line with our hypotheses.

With regards to the overall effect of XAI on AB, due to the low sample size in the pre-test our results are not significant, we can still see some early indications. Especially, commission errors are higher in the XAI setting ($M_{XAI}$ = 21%) than in the AI setting ($M_{AI}$ = 10%). While we cannot explicitly test our hypotheses, we still can identify a tendency that XAI increases AB (**H5**).

|  | AI | | XAI | | |
|---|---|---|---|---|---|
|  | Mean | SD | Mean | SD | P-value |
| **Negative consultations** | 11% | 17% | 17% | 19% | 0.19 (Mann-Whitney U test) |
| **Omission error** | 13% | 22% | 12% | 27% | 0.37 (T-test) |
| **Commission error** | 10% | 21% | 21% | 30% | **0.09** (T-test) |
| **Trust** | 4.44 | 0.84 | 4.37 | 0.93 | 0.44 (T-test) |
| **Mental model goal** | 5.1 | 1.1 | 4.73 | 1.05 | **0.07** (T-test) |

*Table 2.    Differences in the mean of both treatments.*

Additionally, we evaluated the influence of our demographic variables (age, gender, AI experience) on negative consultations by calculating Spearman correlations with no significant results.

For our main study, we plan to study our research model by conducting a structured equation model. We conduct a power analysis to calculate the necessary sample size. Since AB, and consequently negative consultations, is our dependent variable, we calculate the power analysis based on negative consultations in the pre-test. Based on the calculated effect size of 0.29 and a target power of 0.95, a minimum of 257 participants per treatment will be required.

## 6 Discussion and Expected Contribution

In this article, we describe a first pre-test regarding the influence of XAI on automation bias, an over-reliance on AI. To investigate the effect in-depth, we first derive a research model, including hypotheses, and then conduct an online experiment. In the online experiment, we let participants classify hotel reviews as fake and genuine. The results of our study indicate that explanations of AI increase automation bias.

Studies have shown that, in particular cases, significant differences exist between the occurrence of commission and omission errors (Lyell and Coiera 2017). In line with these studies, we find differences between AI and XAI commission errors (still not significant), but no differences in terms of omission errors. One interpretation would be that humans are more likely to classify a review erroneous as fake, i.e., humans tend to believe the AI more if the recommendation detects a state of interest.

Although certainly effects of explanation on automation bias depend on the domain and type of explanations, our results illustrate that XAI may increase over-reliance on AI advice. While one could argue that this is because of confusing state-of-the-art explanations , which is indicated by the qualitative feedback, we believe the underlying reason is far more complex. If the explanations are confusing, it is imminent that humans would have a hard time detecting when the AI is correct. However, it is not intrinsically logical that a confusing explanation increases automation bias. Confusing explanations could also lead to an aversion. Therefore, the underlying mediators of automation bias need to be rigorously examined to understand the relationship between explanations and automation bias.

In terms of the mediators of automation bias in our current study, we do not find any differences between trust in the AI and XAI setting within our limited pre-test. However, we find a non-significant difference in terms of MM. Participants in the AI treatment have reported a higher MM goal. This could mean that humans find it challenging to build a MM of the AI, due to the confusing explanations. Additionally, we conducted a correlation analysis to provide first insights on how XAI influences automation bias. We find no significant results but an indication of the direction of the effects. Trust seems to increase automation bias while a positively influenced MM seems to decrease automation bias which is both in line with our hypotheses. Thus the negative effect of XAI on automation bias in our study could be caused by a negative influence of XAI on the human MM.

In future work, we plan to analyze the influence of explanations on automation bias in-depth. More specifically, in our main study, we plan to investigate the mediation effects more in depth by conducting mediaton analysis (MacKinnon et al. 2007) and structural equation modeling (Ullman and Bentler 2012). By analyzing the mediators of automation bias, we aim to develop further insights into automation bias and thereby create the best possible human-AI collaboration. Despite being a first pre-test, we already contribute to the body of knowledge. We derive a research model and an experimental design to investigate the influence of XAI on automation bias and report as well as discuss first preliminary results. With the commoditization of AI, automation bias will become a serious issue that needs to be analyzed. It must be emphasized that currently explanations of AI always have a positive connotation and are even anchored in law, such as in the European right of explainability (Bauer et al. 2021). Our research indicates that explanations could lead to over-reliance and thus have serious consequences.

## References

Adadi, A., and Berrada, M. 2018. "Peeking Inside the Black-Box: A Survey on Explainable Artificial Intelligence (XAI)," *IEEE Access* (6), IEEE, pp. 52138–52160. (https://doi.org/10.1109/ACCESS.2018.2870052).

Alberdi, E., Povyakalo, A., Strigini, L., and Ayton, P. 2004. "Effects of Incorrect Computer-Aided Detection (CAD) Output on Human Decision-Making in Mammography," *Academic Radiology* (11:8), pp. 909–918. (https://doi.org/10.1016/j.acra.2004.05.012).

Arora, A. 2020. "Conceptualising Artificial Intelligence as a Digital Healthcare Innovation: An Introductory Review," *Medical Devices: Evidence and Research* (13), pp. 223–230. (https://doi.org/10.2147/MDER.S262590).

Bailey, N. R., and Scerbo, M. W. 2007. "Automation-Induced Complacency for Monitoring Highly Reliable Systems: The Role of Task Complexity, System Experience, and Operator Trust," *Theoretical Issues in Ergonomics Science* (8:4), Taylor & Francis, pp. 321–348.

Bansal, G., Nushi, B., Kamar, E., Lasecki, W. S., Weld, D. S., and Horvitz, E. 2019. "Beyond Accuracy: The Role of Mental Models in Human-Ai Team Performance," in *Proceedings of the AAAI Conference on Human Computation and Crowdsourcing* (Vol. 7), pp. 2–11.

Bansal, G., Tongshuang, W. U., Zhou, J., Raymond, F. O. K., Nushi, B., Kamar, E., Ribeiro, M. T., and Weld, D. S. 2020. "Does the Whole Exceed Its Parts? The Effect of AI Explanations on Complementary Team Performance," *ArXiv* (1:1), pp. 1–26.

Bansal, G., Wu, T., Zhou, J., Fok, R., Nushi, B., Kamar, E., Ribeiro, M. T., and Weld, D. 2021. "Does the Whole Exceed Its Parts? The Effect of Ai Explanations on Complementary Team Performance," in *Proceedings of the 2021 CHI Conference on Human Factors in Computing Systems*, pp. 1–16.

Bauer, K., Hinz, O., van der Aalst, W., and Weinhardt, C. 2021. "Expl(AI)n It to Me – Explainable AI and Information Systems Research," *Business and Information Systems Engineering* (63:2), Springer Fachmedien Wiesbaden, pp. 79–82. (https://doi.org/10.1007/s12599-021-00683-2).

Braga, A., and Logan, R. K. 2017. "The Emperor of Strong AI Has No Clothes: Limits to Artificial Intelligence," *Information* (8:4), Multidisciplinary Digital Publishing Institute, p. 156.

Briscoe, E., and Feldman, J. 2011. "Conceptual Complexity and the Bias/Variance Tradeoff," *Cognition* (118:1), Elsevier, pp. 2–16.

Buçinca, Z., Malaya, M. B., and Gajos, K. Z. 2021. "To Trust or to Think: Cognitive Forcing Functions Can Reduce Overreliance on AI in AI-Assisted Decision-Making," *Proceedings of the ACM on Human-Computer Interaction* (5:CSCW1), ACM New York, NY, USA, pp. 1–21.

Bussone, A., Stumpf, S., and O'Sullivan, D. 2015. "The Role of Explanations on Trust and Reliance in Clinical Decision Support Systems," *Proceedings - 2015 IEEE International Conference on Healthcare Informatics, ICHI 2015*, IEEE, pp. 160–169. (https://doi.org/10.1109/ICHI.2015.26).

Challen, R., Denny, J., Pitt, M., Gompels, L., Edwards, T., and Tsaneva-Atanasova, K. 2019. "Artificial Intelligence, Bias and Clinical Safety," *BMJ Quality and Safety* (28:3), pp. 231–237. (https://doi.org/10.1136/bmjqs-2018-008370).

Craik, K. J. W. 1943. "Physiology of Colour Vision," *Nature* (151:3843), Nature Publishing Group, pp. 727–728.

D'Amour, A., Heller, K., Moldovan, D., Adlam, B., Alipanahi, B., Beutel, A., Chen, C., Deaton, J., Eisenstein, J., and Hoffman, M. D. 2020. "Underspecification Presents Challenges for Credibility in Modern Machine Learning," *ArXiv Preprint ArXiv:2011.03395*.

Dellermann, D., Ebel, P., Söllner, M., and Leimeister, J. M. 2019. "Hybrid Intelligence," *Business and Information Systems Engineering* (61:5), pp. 637–643. (https://doi.org/10.1007/s12599-019-00595-2).

Dzindolet, M. T., Peterson, S. A., Pomranky, R. A., Pierce, L. G., and Beck, H. P. 2003. "The Role of Trust in Automation Reliance," *International Journal of Human Computer Studies* (58:6), pp. 697–718. (https://doi.org/10.1016/S1071-5819(03)00038-7).

Gefen, D., Karahanna, E., and Straub, D. W. 2003. "Trust and TAM in Online Shopping: An Integrated Model," *MIS Quarterly*, JSTOR, pp. 51–90.

Goddard, K., Roudsari, A., and Wyatt, J. C. 2012. "Automation Bias: A Systematic Review of Frequency, Effect Mediators, and Mitigators," *Journal of the American Medical Informatics Association* (19:1), pp. 121–127. (https://doi.org/10.1136/amiajnl-2011-000089).

Gregor, S., and Benbasat, I. 1999. "Explanations from Intelligent Systems: Theoretical Foundations and Implications for Practice," *MIS Quarterly: Management Information Systems* (23:4), pp. 497–530. (https://doi.org/10.2307/249487).

Grosz, B. 2012. "What Question Would Turing Pose Today?," *AI Magazine* (33:4), p. 73.

Hemmer, P., Schemmer, M., Vössing, M., and Kühl, N. 2021. "Human-AI Complementarity in Hybrid Intelligence Systems: A Structured Literature Review," *PACIS 2021 Proceedings*.

Hoffman, R. R., Mueller, S. T., Klein, G., and Litman, J. 2018. "Metrics for Explainable AI: Challenges and Prospects," *ArXiv Preprint ArXiv:1812.04608*.

Jussupow, E., Spohrer, K., Heinzl, A., and Gawlitza, J. 2021. "Augmenting Medical Diagnosis Decisions? An Investigation into Physicians' Decision-Making Process with Artificial Intelligence," *Information Systems Research*, INFORMS.

Khasawneh, R., and Kornreich, R. 2014. "Explaining Data-Driven Document Classifications," *MIS Quarterly: Management Information Systems* (3:4), pp. 781–791.

Koehler, D. J. 1991. "Explanation, Imagination, and Confidence in Judgment.," *Psychological Bulletin* (110:3), American Psychological Association, p. 499.

Kulesza, T., Stumpf, S., Burnett, M., and Kwan, I. 2012. "Tell Me More? The Effects of Mental Model Soundness on Personalizing an Intelligent Agent," in *Proceedings of the SIGCHI Conference on Human Factors in Computing Systems*, pp. 1–10.

Lai, V., Liu, H., and Tan, C. 2020. "Why Is 'Chicago' Deceptive? Towards Building Model-Driven Tutorials for Humans," *ArXiv*, pp. 1–13.

Lai, V., and Tan, C. 2019. "On Human Predictions with Explanations and Predictions of Machine Learning Models: A Case Study on Deception Detection," *FAT\* 2019 - Proceedings of the 2019 Conference on Fairness, Accountability, and Transparency*, pp. 29–38. (https://doi.org/10.1145/3287560.3287590).

Lecun, Y., Bengio, Y., and Hinton, G. 2015. "Deep Learning," *Nature* (521:7553), pp. 436–444. (https://doi.org/10.1038/nature14539).

Lee, J. D., and See, K. A. 2004. "Trust in Automation: Designing for Appropriate Reliance," *Human Factors* (46:1), SAGE Publications Sage UK: London, England, pp. 50–80.

Van Lent, M., Fisher, W., and Mancuso, M. 2004. "An Explainable Artificial Intelligence System for Small-Unit Tactical Behavior," in *Proceedings of the National Conference on Artificial Intelligence*, Menlo Park, CA; Cambridge, MA; London; AAAI Press; MIT Press; 1999, pp. 900–907.

Liel, Y., and Zalmanson, L. 2020. "What If an AI Told You That 2 + 2 Is 5 ? Conformity to Algorithmic Recommendations," *ICIS 2020*, pp. 0–9.

Liu, H., Lai, V., and Tan, C. 2021. *Understanding the Effect of Out-of-Distribution Examples and Interactive Explanations on Human-AI Decision Making*, pp. 1–42. (http://arxiv.org/abs/2101.05303).

Loewenstein, G. 1999. "Experimental Economics from the Vantage-Point of Behavioural Economics," *The Economic Journal* (109:453), JSTOR, pp. F25–F34.

Lyell, D., and Coiera, E. 2017. "Automation Bias and Verification Complexity: A Systematic Review," *Journal of the American Medical Informatics Association* (24:2), pp. 423–431. (https://doi.org/10.1093/jamia/ocw105).

MacKinnon, D. P., Fairchild, A. J., and Fritz, M. S. 2007. "Mediation Analysis," *Annu. Rev. Psychol.* (58), Annual Reviews, pp. 593–614.

Meske, C., Bunde, E., Schneider, J., and Gersch, M. 2020. "Explainable Artificial Intelligence: Objectives, Stakeholders, and Future Research Opportunities," *Information Systems Management* (00:00), Taylor & Francis, pp. 1–11. (https://doi.org/10.1080/10580530.2020.1849465).

Mosier, K. L., and Skitka, L. J. 1999. "Automation Use and Automation Bias," in *Proceedings of the Human Factors and Ergonomics Society Annual Meeting* (Vol. 43), SAGE Publications Sage CA: Los Angeles, CA, pp. 344–348.

Ott, M., Cardie, C., and Hancock, J. T. 2013. "Negative Deceptive Opinion Spam," in *Proceedings of the 2013 Conference of the North American Chapter of the Association for Computational Linguistics: Human Language Technologies*, pp. 497–501.

Ott, M., Choi, Y., Cardie, C., and Hancock, J. T. 2011. "Finding Deceptive Opinion Spam by Any Stretch of the Imagination," *ACL-HLT 2011 - Proceedings of the 49th Annual Meeting of the Association for Computational Linguistics: Human Language Technologies* (1), pp. 309–319.

Ribeiro, M. T., Singh, S., and Guestrin, C. 2016. "'Why Should i Trust You?' Explaining the Predictions of Any Classifier," *Proceedings of the ACM SIGKDD International Conference on Knowledge Discovery and Data Mining* (13-17-Augu), pp. 1135–1144. (https://doi.org/10.1145/2939672.2939778).

Rouse, W. B., and Morris, N. M. 1986. "On Looking into the Black Box: Prospects and Limits in the Search for Mental Models.," *Psychological Bulletin* (100:3), American Psychological Association, p. 349.

Santiago, T. 2019. "AI Bias: How Does AI Influence The Executive Function Of Business Leaders?," *Muma Business Review* (3), pp. 181–192.

Savage-Knepshield, P. A. 2001. *Mental Models: Issues in Construction, Congruency, and Cognition*, Rutgers The State University of New Jersey-New Brunswick.

Schemmer, M., Hemmer, P., Kühl, N., Benz, C., and Satzger, G. 2022. "Should I Follow AI-Based Advice? Measuring Appropriate Reliance in Human-AI Decision-Making," in *ACM CHI 2022 Workshop on Trust and Reliance in AI-Human Teams (TrAIt)*.

Skitka, L. J., Mosier, K. L., and Burdick, M. 1999. "Does Automation Bias Decision-Making?," *International Journal of Human Computer Studies* (51:5), pp. 991–1006. (https://doi.org/10.1006/ijhc.1999.0252).

Ullman, J. B., and Bentler, P. M. 2012. "Structural Equation Modeling," *Handbook of Psychology, Second Edition* (2), Wiley Online Library.

Vitharana, P., Fatemeh, M., Zahedi, S. B., and Jain, H. K. 2016. "Enhancing Analysts' Mental Models for Improving Requirements Elicitation: A Two-Stage Theoretical Framework," *Journal of the Association for Information Systems* (17:12), pp. 804–840.

Wanner, J., Herm, L.-V., Heinrich, K., Janiesch, C., and Zschech, P. 2020. "White, Grey, Black: Effects of XAI Augmentation on the Confidence in AI-Based Decision Support Systems," *Proceedings of the Forty-First International Conference on Information Systems*, pp. 0–9.

Weitz, K., Schiller, D., Schlagowski, R., Huber, T., and André, E. 2019. "' Do You Trust Me?' Increasing User-Trust by Integrating Virtual Agents in Explainable AI Interaction Design," in *Proceedings of the 19th ACM International Conference on Intelligent Virtual Agents*, pp. 7–9.

Weld, D. S., and Bansal, G. 2019. "The Challenge of Crafting Intelligible Intelligence," *Communications of the ACM* (62:6), ACM New York, NY, USA, pp. 70–79.

Zhang, Y., Vera Liao, Q., and Bellamy, R. K. E. 2020. "Efect of Confidence and Explanation on Accuracy and Trust Calibration in AI-Assisted Decision Making," *FAT\* 2020 - Proceedings of the 2020 Conference on Fairness, Accountability, and Transparency*, pp. 295–305. (https://doi.org/10.1145/3351095.3372852).